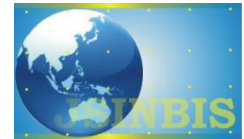

# Prototype Software Monitoring Sarana dan Prasarana Perguruan Tinggi

Leon Andretti Abdillah[*], Linda Atika, Kurniawan, Fitri Purwaningtias

Information Systems Department, Universitas Bina Darma

**Abstract**

This study aims to facilitate the management system of monitoring infrastructure program of university facilities and infrastructure, through software engineering technology approach as an effort to improve productivity and quality of monitoring process become more efficient and effective. The software in this research is built in a systematic and organized approach to monitoring infrastructure of facilities and infrastructure using appropriate tools and techniques. Through this research, universities are expected to be able to develop the necessary quality measures to support the process of planning and controlling infrastructure infrastructure monitoring. The research was conducted using survey method, development of monitoring management and software. Up to the design stage of this program prototype, research has produced a special picture of the software requirements to be built in the next year. Software development process starting from the analysis phase of system and software requirements, designing data structures up to the architecture stage of the program has produced a list of needs/requirements, the design of program prototype contained in the design of input/output for the monitoring process facilities and infrastructure.

**Abstrak**

Penelitian ini bertujuan untuk memfasilitasi sistem manajemen pemantauan program infrastruktur sarana dan prasarana universitas, melalui pendekatan teknologi rekayasa perangkat lunak sebagai upaya meningkatkan produktivitas dan kualitas proses pemantauan menjadi lebih efisien dan efektif. Perangkat lunak dalam penelitian ini dibangun dengan pendekatan sistematis dan terorganisasi untuk memantau infrastruktur fasilitas dan infrastruktur menggunakan alat dan teknik yang tepat. Melalui penelitian ini, universitas diharapkan dapat mengembangkan langkah-langkah kualitas yang diperlukan untuk mendukung proses perencanaan dan pengendalian pemantauan infrastruktur infrastruktur. Penelitian ini dilakukan dengan menggunakan metode survei, pengembangan manajemen pemantauan dan perangkat lunak. Sampai tahap desain prototipe program ini, penelitian telah menghasilkan gambaran khusus dari persyaratan perangkat lunak yang akan dibangun pada tahun depan. Proses pengembangan perangkat lunak mulai dari tahap analisis sistem dan kebutuhan perangkat lunak, perancangan struktur data hingga tahap arsitektur program telah menghasilkan daftar kebutuhan / persyaratan, perancangan prototipe program yang terdapat dalam perancangan input / output untuk fasilitas dan infrastruktur proses pemantauan.

## 1. Pendahuluan

Salah satu dampak dari globalisasi saat ini adalah meluasnya penggunaan teknologi informasi (TI) dalam mendukung berbagai kegiatan. TI telah digunakan secara luas dalam banyak aspek kehidupan kita sehari-hari (Abdillah, 2016). TI memiliki kemampuan untuk mengolah data menjadi informasi yang berguna dengan cepat dan akurat. Apalagi dengan munculnya teknologi internet, semua kegiatan dapat dilakukan menggunakan halaman web. Per Januari 2018 (Kemp, 2018), total populasi total dunia sekitar 7,476 miliar, dimana 53% dari mereka adalah pengguna internet. Data ini menginformasikan kepada kita bahwa internet menjadi media utama bagi orang untuk berkomunikasi satu sama lain, mulai dari perbankan, pemerintah, bahkan ke lingkungan pendidikan.

Pendidikan merupakan infrastruktur utama (Abdillah, 2013) untuk mempromosikan suatu bangsa dan meningkatkan daya saing mereka di era globalisasi. Universitas diminta untuk menyediakan sumber daya manusia yang berkualitas, sistem pendanaan, dan fasilitas dan infrastruktur untuk mendukung operasionalisasi kegiatan akademik. Artikel ini merupakan kelanjutan dari publikasi sebelumnya (Atika *et al.*, 2017). Sebagai aset yang sangat penting, sumber daya manusia yang ada perlu ditingkatkan pengelolaannya dengan lebih baik, terutama yang berkaitan dengan sarana dan prasarana (Abdillah *et al.*, 2007).

---

*) Penulis korespondensi: leon.abdillah@yahoo.com

Fasilitas dan Infrastruktur adalah salah satu komponen yang harus ada di universitas. Berdsaran Permendiknas No. 19 Tahun 2005 tentang Standar Pendidikan Nasional Bab VII Standar Sarana dan Prasarana Pasal 42 paragraf 1 (Presiden Republik Indonesia, 2005): "Setiap satuan pendidikan wajib memiliki sarana yang meliputi perabot, peralatan pendidikan, media pendidikan, buku dan sumber belajar lainnya, bahan habis pakai, serta perlengkapan lain yang diperlukan untuk menunjang proses pembelajaran yang teratur dan berkelanjutan.". Kemudian paragraf 2 mengatakan "Setiap satuan pendidikan wajib memiliki prasarana yang meliputi lahan, ruang kelas, ruang pimpinan satuan pendidikan, ruang pendidik, ruang tata usaha, ruang perpustakaan, ruang laboratorium, ruang bengkel kerja, ruang unit produksi, ruang kantin, instalasi daya dan jasa, tempat berolahraga, tempat beribadah, tempat bermain, tempat berkreasi, dan ruang/tempat lain yang diperlukan untuk menunjang proses pembelajaran yang teratur dan berkelanjutan."

Sarana dan prasarana merupakan bagian penting dalam pemenuhan proses pencapaian pembelajaran di perguruan tinggi. Agar kualitas sarana dan prasarana terjamin maka diperlukan proses pemantauan sarana dan prasarana. Penggunaan sarana dan prasarana banyak yang belum menggunakan prosedur operasi standar dan data dalam bentuk digital.

Saat ini, sarana dan prasarana pendidikan tinggi di Indonesia dalam mendukung proses pembelajaran berkualitas masih belum memadai. Fenomena ini tidak hanya terjadi di daerah terpencil tetapi di kota-kota besar sekalipun kita masih menemukan, misalnya masih ada ruang kelas yang tidak layak sebagai tempat belajar kuliah (Yuliawati, 2012). Keberhasilan perangkat lunak yang dibangun dilihat berdasarkan apakah perangkat lunak bekerja atau tidak pada proses bisnis yang sedang berjalan (Bachtiar *et al.,* 2013).

Di universitas-universitas di Indonesia, khususnya provinsi Sumatera Selatan belum banyak menggunakan perangkat lunak dalam memonitor sarana dan prasarana. Salah satu universitas di kota Palembang, salah satu contohnya adalah Universitas Bina Darma. Saat ini, Universitas Bina Darma dalam mengelola sarana dan prasarana sudah memiliki sistem informasi yang disebut SIMPERANG (Sistem Informasi Pengadaan) untuk proses pengajuan barang, tetapi dalam hal sarana dan prasarana pemantauan, jika ditemukan kerusakan atau ketidaksesuaian sarana dan prasarana maka unit kerja harus mengisi formulir temuan infrastruktur yang diserahkan kepada Biro Penjaminan Mutu (BPM) yang berfungsi sebagai unit pemantauan infrastruktur. Temuan akan disurvei dan ditindaklanjuti melalui lisan atau mengirim surat dari biro administrasi, atau kepada para pemangku kepentingan yang akan melakukan perbaikan tersebut.

BPM juga harus mengisi formulir pemantauan fasilitas dan infrastruktur. Setelah itu, jika kerusakan atau ketidaksesuaian sarana dan prasarana telah diperbaiki maka Biro Penjaminan Mutu harus menemukan temuan infrastruktur dalam arsip temuan infrastruktur dan mengisi tanggal temuan setelah proses perbaikan selesai. Secara berkala, Biro Jaminan Mutu juga akan membuat laporan untuk diberikan kepada yayasan.

Proses monitoring ini hingga proses pembuatan laporan yang memakan waktu lama dan tidak efisien karena tidak ada aplikasi yang mendukung proses monitoring sarana dan prasarana yang ada di perguruan tinggi. Jika proses monitoring dan pelaporan tidak terlalu bertele-tele dan sarana dan prasarana dalam kondisi baik atau siap pakai sehingga membuat siswa nyaman, karyawan, dosen dan pemimpin maka akan membantu proses pembelajaran menjadi lebih efektif.

Persiapan sistem manajemen pemantauan dimulai dari data survei ke beberapa universitas di Palembang. Yang kedua menganalisis masalah yang ada dalam proses pemantauan barang, dan ketiga untuk melakukan analisis pustaka sehingga sistem informasi sarana dan prasarana pemantauan yang akan dikembangkan sesuai dengan kebutuhan pengguna.

Hasil survei menunjukkan bahwa: 1) setiap universitas belum memiliki perangkat lunak yang membantu bagian pemantauan untuk memantau kondisi sarana dan prasarana, 2) Registrasi monitoring barang masih dilakukan secara manual, dimana sarana dan prasarana harus memeriksa file yang masuk untuk melakukan monitoring dan kemudian memutuskan apakah layak untuk melakukan perbaikan barang sehingga informasi yang diterima lambat, 3) Proses penyaringan barang masih sulit untuk dilihat misalnya perbaikan terbaru yang telah dilakukan seperti pemeliharaan AC yang dilakukan secara berkala, peralatan IT yang memiliki masa garansi tetapi sulit untuk melacak apakah peralatan IT sudah keluar dari masa garansi, 4) Proses pelaporan dan temuan masih dilakukan secara manual, sehingga pihak pelapor harus mengirim surat ke fasilitas dan infrastruktur untuk ditindaklanjuti.

Fokus pada makalah ini adalah bagaimana mengembangkan perangkat lunak pemantauan infrastruktur prototipe di lembaga pendidikan tinggi. Secara umum, penelitian ini bertujuan untuk merekayasa proses pemantauan fasilitas dan infrastruktur universitas untuk meningkatkan kualitas data dan informasi kondisi barang dengan cepat, tepat waktu dan akurat untuk mendukung proses pengambilan keputusan. Adapun manfaat dari penelitian ini: 1) meningkatkan produktivitas dan kualitas proses pemantauan fasilitas dan infrastruktur di universitas, dan 2) untuk membantu dan mendukung unit kerja yang menerapkan fasilitas pemantauan dan infrastruktur pendidikan tinggi.

Sisa artikel ini diatur dalam 3 (tiga) bagian lagi. Bagian II menjelaskan aliran penelitian tentang bagaimana penelitian dilakukan sebagai metode penelitian. Pada bagian III, penulis menampilkan beberapa contoh antarmuka sebagai hasil dan diikuti oleh bagian diskusi. Akhirnya, penulis menulis beberapa catatan sebagai kesimpulan dan arahan untuk pekerjaan selanjutnya di bagian terakhir, Bagian IV.

**2. Kerangka Teori**

Pada penelitian ini Penelitian ini merupakan penelitian deskriptif kualitatif dan penelitian tindakan dengan melakukan observasi, wawancara mendalam, dan dokumen.

*2.1. Prototipe Software*

Prototipe merupakan salah satu bentuk pengembangan perangkat lunak (*software*). Sebelum suatu *software* dibangun dengan menggunakan bahasa pemrograman tertentu perlu dibuatkan protitipe yang dapat dilihat oleh berbagai pihak terkait. Dengan pendekatan protitipe kebutuhan setiap pihak akan dengan mudah terakomodir.

Dalam pembuatan prototipe, tim pengembang akan dengan lebih cepat menampilkan perkiraan tampilan dari suatu *software*. Sehingga apabila ditemukan kekurangan-kekurangan akan dengan mudah diketahui dan segera diperbaiki. Pada model prototipe ini, dapat dinilai secara iteratif oleh pemangku kepentingan (*stakeholders*) dalam upaya mengidentifikasi atau memperkuat persyaratan perangkat lunak.

Prototipe akan memberi beberapa hal positif (Kroll dan Kruchten, 2003) yakni 1) Sesuatu yang lebih konkret untuk ditampilkan, 2) Pengembang dapat memvalidasi bahwa mereka benar-benar memiliki semua bagian yang harus dilakukan, 3) Lebih banyak informasi untuk membuat rencana yang jauh lebih baik dan jadwal yang lebih rinci, 4) Titik awal untuk membangun hal yang nyata (*full coding*), dengan kesempatan untuk menghapus semua kesalahan-kesalahan yang ada.

Alasan umum untuk membuat prototipe termasuk membuktikan kelayakan teknis dan memvalidasi kegunaan antarmuka pengguna (Braude dan Bernstein, 2016). Tren pengembangan perangkat lunak dengan cepat telah mendorong banyak pihak pengembang untuk menggunakan protitipe.

*2.2. Sarana dan Prasarana Perguruan Tinggi*

Prasarana akademik adalah perangkat penunjang utama suatu proses atau usaha pendidikan agar tujuan pendidikan tercapai. Sementara, sarana adalah segala sesuatu yang dapat dipakai sebagai alat/media dalam mencapai maksud atau tujuan.

Selain itu, sarana dan prasarana juga merupakan salah satu Standar Pendidikan Nasional Tahun 2014 (Menristekdikti, 2014) dan akan dinilai oleh Badan Akreditasi Nasional Perguruan Tinggi (BAN-PT). Standar pembelajaran harus terdiri dari: 1) Furniture, 2) Peralatan pendidikan, 3) Media pendidikan, 4) Buku, buku elektronik, dan repositori, 5) fasilitas teknologi informasi dan komunikasi, 6) Instrumentasi eksperimental, 7) Fasilitas olahraga , 8) Fasilitas seni, 9) Fasilitas fasilitas umum, 10) Perlengkapan; dan 11) Fasilitas pemeliharaan, keselamatan dan keamanan.

Menurut Peraturan Menteri Pendidikan dan Kebudayaan tahun 2013 tentang Standar Pendidikan Nasional Pendidikan Tinggi Pasal 38, disebutkan bahwa: 1) Setiap perguruan tinggi wajib memiliki sarana untuk melaksanakan Tridharma perguruan tinggi yang meliputi: a) perabot, b) peralatan pembelajaran, c) media pembelajaran, d) buku dan sumber belajar lain, e) teknologi informasi dan komunikasi, f) bahan habis pakai, g) perlengkapan lain yang diperlukan, dan 2) Setiap perguruan tinggi wajib memenuhi prasarana untuk melaksanakan Tridharma perguruan tinggi yang meliputi: a) lahan, b) ruang kelas, c) ruang pimpinan perguruan tinggi, d) ruang perpustakaan, e) ruang laboratorium, f) ruang bengkel kerja, g) ruang unit produksi, h) ruang kantin, i) tempat berolahraga.

*2.3. Penelitian Terkait*

Beberapa penelitian terkait sebelumnya termasuk: 1) *Development of a Remote Building Monitoring System* (Olken *et al.*, 1998), 2) *Software development for infrastructure* (Stroustrup, 2012), dan 3) Evaluasi dan Perancangan Sistem Informasi Pemeliharaan Sarana Prasarana di Fakultas Ilmu Komputer Universitas Brawijaya (Widianto, 2016).

**3. Metode**

Penelitian ini merupakan penelitian deskriptif kualitatif dan penelitian tindakan dengan melakukan observasi, wawancara mendalam, dan dokumen. Metode ini dikombinasikan untuk menggambarkan semua fakta yang terkait dengan peningkatan pemantauan fasilitas dan infrastruktur di universitas.

Jenis dan sumber data yang digunakan dalam penelitian ini adalah: 1) data primer adalah data yang dapat berasal dari responden dan sumber informan pertama yaitu individu atau individu seperti hasil kuesioner dan wawancara yang dilakukan oleh peneliti. Data primer berupa rekaman hasil wawancara, hasil observasi ke lapangan langsung berupa catatan tentang situasi dan kejadian, dan data tentang informan, dan 2) data sekunder yaitu data primer yang telah diproses lebih lanjut dan disajikan oleh pengumpul data primer pihak atau yang lain seperti dalam tabel, diagram dan foto dan dokumen yang relevan. Data ini digunakan untuk mendukung informasi utama yang diperoleh baik dari dokumen, atau dari pengamatan langsung ke lapangan.

Informan dalam penelitian ini adalah universitas dalam memantau sarana dan prasarana. Informasi dikumpulkan dengan menggunakan *brainstorming*. *Brainstorming* adalah alat yang sangat bagus untuk mendiskusikan masalah (Abdillah *et al.*, 2018). Setelah semua informasi yang dibutuhkan dikumpulkan, maka semua itu akan diproses untuk mengembangkan perangkat lunak dalam fasilitas dan infrastruktur pendidikan tinggi.

Teknik pengolahan dan analisis data penelitian direncanakan dimulai dengan melakukan kajian pustaka dalam bentuk konsep teori dan hasil penelitian yang relevan. Hasil penelitian ini membentuk dasar untuk membangun manajemen yang efektif dan efisien dan sistem pemantauan untuk fasilitas dan infrastruktur.

Untuk mengeksplorasi masalah, akan diamati ke lapangan untuk melihat langsung kondisi sarana dan prasarana yang ada. Data awal akan dikumpulkan untuk menggambarkan faktor-faktor yang relevan dengan apa yang akan ditingkatkan atau dikembangkan. Hasil dari tahap ini adalah proses pemantauan sistem pengelolaan sarana dan prasarana sesuai dengan kondisi aktual perguruan tinggi.

Berdasarkan model teoritis dari optimasi yang dihasilkan, maka dilakukan langkah analisis dan perancangan sistem. Pada tahap ini akan dilakukan analisis data hasil survei dan observasi serta perancangan sarana dan prasarana sistem pemantauan. Selanjutnya, lanjutkan dengan proses pengembangan perangkat lunak. Serangkaian uji coba akan dilakukan, setelah prototipe dinilai telah mencapai hasil yang diinginkan dan menjawab masalah penelitian akan menghasilkan prototipe pemantauan fasilitas dan infrastruktur.

*Diagram fishbone* adalah diagram sebab dan akibat (Mahanti dan Antony, 2005), juga disebut sebagai "Ishikawa diagram". *Diagram fishbone* memberikan bentuk visual seperti ikan untuk ringkasan hambatan untuk fasilitas dan infrastruktur yang memadai. Diagram ini (Berry dan Dahl, 2000) digunakan sebagai alat untuk memetakan faktor-faktor yang diduga mempengaruhi masalah atau hasil yang diinginkan (lihat gambar 1).

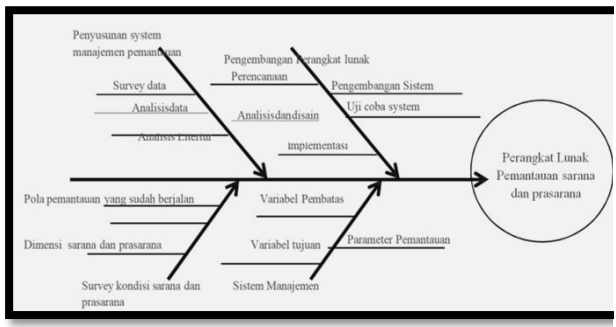

Gambar 1. Diagram Aliran Penelitian Fishbone.

Prototipe perangkat lunak dikembangkan dengan mengikuti langkah-langkah dalam metode prototipe atau paradigma. Paradigma ini telah direkomendasikan sebagai pendekatan pengembangan sistem alternatif (Guimarae dan Saraph, 1991). Paradigma *prototyping* (Pressman, 2010) terdiri atas: 1) *Communication*, 2) *Quick plan*, 3) *Modelling Quick design*, 4) *Construction of prototype*, dan 5) *Delpoyment Delivery & Feedback*.

Prototipe cepat dapat lebih baik dalam kecepatan, akurasi, dan kurang melelahkan daripada metode konvensional menciptakan model (Neeley *et al.*, 2013). Prototipe adalah versi awal (Sommerville, 2009) dari sistem perangkat lunak yang digunakan untuk memperkenalkan konsep, menguji opsi desain, dan mencari tahu lebih lanjut tentang masalah dan kemungkinan solusinya (lihat gambar 2).

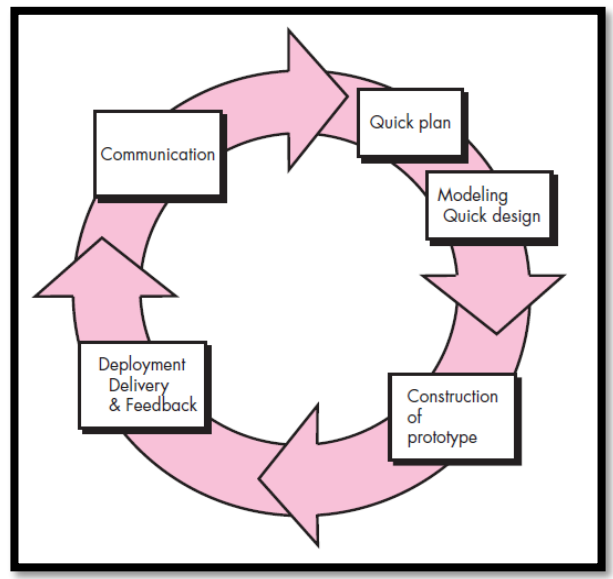

Gambar 2. Paradigma Prototyping.

Prototipe ini ditulis dengan menggunakan PHP sebagai salah satu bahasa pemrograman skrip paling terkenal dalam mengembangkan sebuah situs web. PHP adalah bahasa scripting yang bersifat *general-purpose* (Jibaja, Blackburn, Haghighat, & McKinley, 2011) yang sebagian besar digunakan untuk pengembangan *web* sisi *server*. Untuk database, penulis melibatkan sistem manajemen basis data relasional *open-source* (RDBMS) disebut MySQL. MySQL sangat umum digunakan dengan PHP. MySQL memiliki beberapa kelebihan dan kekuatan (Pore dan Pawar, 2015) sebagai berikut: 1) *replication* (beban kerja dapat dikurangi berat), 2) *sharding*, 3) meningkatkan kinerja operasi pengambilan data, kedewasaan (telah digunakan untuk waktu yang lama), berbagai platform dan bahasa, dan biaya efektif (bebas biaya dan basis data sumber terbuka).

## 3. Hasil dan Pembahasan

Berikut akan diuraikan sejumlah hasil diikuti dengan bahasan terkait prototipe pemantauan sarana dan prasarana di perguruan tinggi.

### 3.1. System Map and Login

Prototipe pemantauan fasilitas dan infrastruktur ini disebut "SIMANTAUBARANG". Secara keseluruhan sistem ini akan terdiri dari 5 (lima) menu utama, yaitu: 1) *Reference*, 2) *Processing*, 3) *Assistant*, 4) *Galerry*, dan 5) *Other*. Lima menu utama, memiliki sejumlah 16 (enam belas) submenu yang dapat dilihat pada gambar 3.

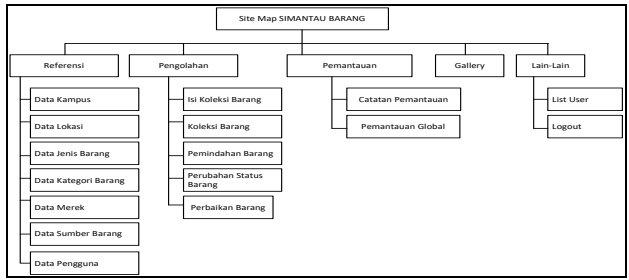

Gambar 3. Peta prototipe sistem pemantauan.

Halaman login ini ditujukan untuk pengguna yang memiliki hak akses yang berbeda. Para pengguna dalam aplikasi ini adalah: 1) Bagian sarana dan prasarana, 2) Unit kerja, dan 3) Pimpinan.

Pengguna akan diminta untuk memasukkan nama pengguna dan kata sandi, jika nama pengguna dan kata sandi sesuai dengan data yang terdapat dalam database, maka sistem akan menampilkan menu utama. Setelah Login, halaman depan yang muncul dalam aplikasi ini berupa data referensi (lihat gambar 4).

Pengguna akan diminta untuk memasukkan nama pengguna dan kata sandi, jika nama pengguna dan kata sandi sesuai dengan data yang terdapat dalam database, maka sistem akan menampilkan menu utama.

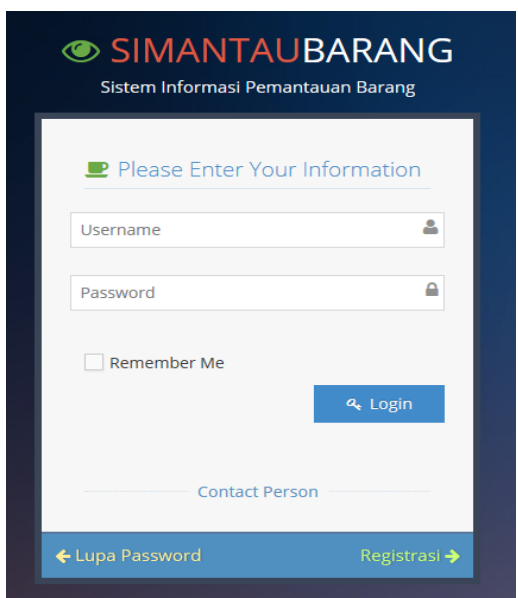

Gambar 4. Login.

### 3.2. Systems Main Menu

Setelah Login, halaman depan yang muncul dalam aplikasi ini berupa data referensi (lihat gambar 5).

Berdasarkan karakteristik pengguna maka dapat dirancang menu yang akan digunakan dalam memantau sarana dan prasarana. Data referensi meliputi (data kampus, lokasi, jenis barang, kategori barang, barang merek, sumber, data pengguna), pemrosesan yang digunakan untuk memproses data barang dengan mengetahui tanggal pembelian, tanggal masa garansi, kondisi, lokasi, sumber barang, dan penanggung jawab barang (lihat gambar 5).

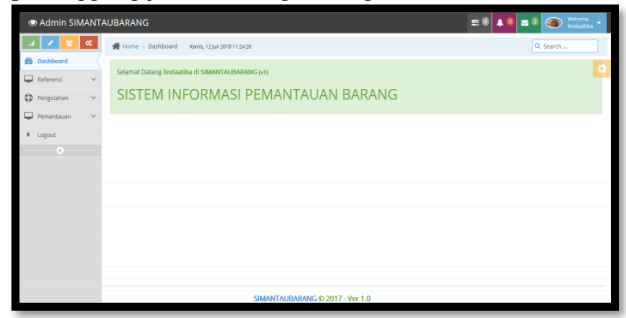

Gambar 5. *Main Menu*.

### 3.3. Reference Campus Data

Setelah memasukkan data referensi ada menu input data kampus, di mana sarana dan prasarana akan menginput data kampus yang ada di Perguruan Tinggi. *Reference* adalah salinan yang dapat memberikan informasi tentang data kampus, lokasi, jenis barang, kategori barang, merek, dan sumber barang. Contoh referensi kampus data terdiri dari ID, Kode kampus, Nama Kampus, dan Alamat Kampus (lihat gambar 6).

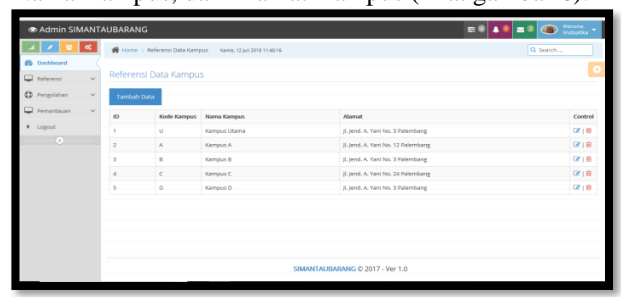

Gambar 6. *Reference campus data*.

### 3.4. Location Setting

Setelah memasukkan data kampus, maka admin akan memasukkan data lokasi ruangan tempat lokasi kampus berada. Misalnya, ruang tipe admin B.201, yang ada di lantai dua dan terletak di Kampus B (lihat gambar 7).

Pada menu input data penerimaan barang, fasilitas dan infrastruktur akan menetralkan data barang masuk, spesifikasi, jenis, merek dan mengunggah foto barang yang masuk.

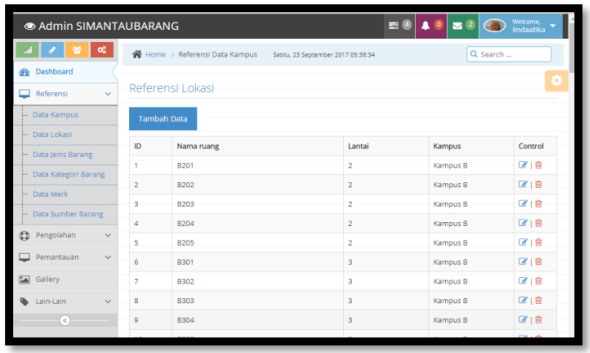

Gambar 7. *Setting location*.

### 3.5. Input Monitoring Data

Proses utama pemantauan dimulai dari memasukkan pemantauan barang, unit kerja akan memasukkan kode barcode, tanggal, nama barang, lokasi, temuan dan rekomendasi (lihat gambar 8).

*Processing* adalah proses bekerja atau bekerja pada data untuk membuatnya lebih sempurna. Submenu yang ada dalam pengolahan adalah isi dari koleksi barang, koleksi barang, transfer barang, perubahan status barang, dan perbaikan barang.

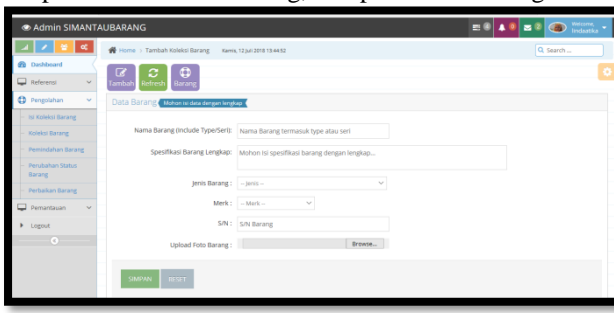

Gambar 8. *Processing*.

### 3.6. Monitoring

*Monitoring* adalah memeriksa penampilan dan aktivitas yang dilakukan. Submenu yang ada dalam pemantauan adalah catatan pemantauan, pemantauan global, barang rusak ringan, barang rusak berat, barang hilang, barang yang disumbangkan, dan berdasarkan lokasi. Contoh sistem yang menampilkan daftar catatan hasil pemantauan barang dapat dilihat pada gambar 9.

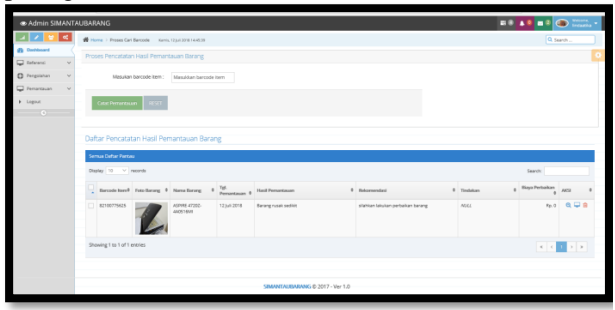

Gambar 9. *Monitoring*.

Admin dapat melakukan pemantauan secara global. Dalam proses *monitoring* admin dapat mengisi nama objek, deskripsi objek pemantauan, temuan, rekomendasi, pemantauan upload dari tampilan depan, tampilan samping, tampilan belakang, dan produk serial. Prototipe sistem juga dapat melihat koleksi item yang rusak ringan, rusak berat, hilang, diberikan, dan melihat koleksi barang berdasarkan lokasi.

### 4. Kesimpulan

Kesimpulan sampai tahap desain prototipe program ini, peneliti telah menghasilkan gambar khusus dari perangkat lunak yang perlu dibangun pada fase berikutnya. Proses pengembangan perangkat lunak yang dimulai dari tahap analisis persyaratan sistem dan perangkat lunak, desain struktur data hingga tahap arsitektur program telah menghasilkan daftar persyaratan, desain prototipe program yang terdapat dalam desain input/output untuk fasilitas proses pemantauan dan infrastruktur. Peneliti menyadari masih banyak kekurangan untuk tahap desain prototipe program ini. Pada fase berikutnya, peneliti bermaksud untuk melakukan pengujian dan evaluasi terhadap pengguna aplikasi SIMANTAU BARANG sehingga nantinya aplikasi dapat dibangun sesuai dengan keinginan pengguna dan bermanfaat untuk Pendidikan Tinggi, terutama dalam memonitor sarana dan prasarana.

SIMANTAUBARANG adalah aplikasi yang digunakan untuk memantau fasilitas dan infrastruktur di universitas dengan berbagai kategori termasuk: 1) Mesin ketik dan Hitung, 2) Alat Reproduksi (Pengganda), 3) Peralatan Penyimpanan Peralatan Ktr, 4) Alat Kantor Lainnya, 5) Peralatan Rumah Tangga, 6) Alat Pembersih, 7) Perangkat Pendingin, 8) Peralatan Dapur, 9) Peralatan Rumah Berlangganan Lainnya, 10) Alat Pemadam Kebakaran, 11) Komputer, 12) Komputer Pribadi, 13) Peralatan Komputer Mainframe, 14) Peralatan Komputer Mini, 15) ) Peralatan Komputer Pribadi, 16) Peralatan Jaringan, 17) Peralatan Studio dan Peralatan Komunikasi, 18) Peralatan Video dan Film Studio, 19) Peralatan Video dan Film Studio A, 20) Peralatan Percetakan.

Untuk tahap selanjutnya, peneliti sedang menguji tahap pengujian untuk mendapatkan perangkat lunak yang bebas kesalahan (*bug*) dan sesuai dengan keinginan pengguna. Pada tahap ini juga akan diuji dan implementasi langsung dengan pengguna dan untuk mengevaluasi penggunaan fasilitas dan infrastruktur perangkat lunak pemantauan. Peneliti akan mendesain aplikasi berbasis web dan berbasis mobile sehingga pengguna dapat dengan mudah menggunakan perangkat lunak SIMANTAU di mana saja dan kapan saja ketika mereka terhubung ke jaringan untuk membantu pengguna dalam memantau fasilitas dan infrastruktur di universitas, terutama di Kota Palembang.

**Ucapan Terima Kasih**

Penelitian ini didukung oleh Direktorat Riset dan Pengabdian Masyarakat (DRPM) Ristek DIKTI [grant number: 03/E/KPT/2018] untuk tahun fiskal 2018 dalam skema "Penelitian Strategis Nasional Institusi" (PSNI).

**Daftar Pustaka**